\documentclass{article}
\usepackage{amsmath}
\usepackage{amssymb}
\usepackage{graphicx}
\begin{document}
\begin{center}
\bfseries
\Large
A MAJOR ORIGIN OF QUANTUMNESS
\normalsize
\end{center}
\begin{center}
Gerd Niestegge
\end{center}
\begin{center}
\footnotesize
https://orcid.org/0000-0002-3405-9356\\
gerd.niestegge@web.de
\end{center}
\begin{abstract}
A system property is considered to be elementary,
when any further refined specification is impossible.
Any two elementary properties of a classical system
mutually exclude each other. However, this does not
hold anymore for a quantum system, where 
mutual exclusivity becomes a special relation.
This orthogonality relation becomes our first axiom.
It is sufficient to define states in an abstract setting
and we then introduce
two simple, but far-reaching further basic axioms for the states,
which mimic the quantum mechanical correspondence 
between the pure states and the rank-one projections.
All examples that we can identify origin from operator algebras
or certain Jordan algebras and require the Gleason theorem.
The identification of further cases
is an unresolved challenge.
If some exist, they might result in promising new mathematical
models for an extended quantum theory. The non-existence
would mean that our extremely basic postulates
bring us unexpectedly close to common quantum theory.
In any case, the distinction between inequality and mutual exclusivity
of elementary system properties appears to be
a major origin of quantumness.
Our basic approach does not encompass the
non-atomic von Neumann algebras of type II and III,
which play a crucial role in quantum field theory,
and we present a modification of the axioms
to include them.\\

\end{abstract}

\section{Introduction}

A system property is considered to be elementary,
when any further refined specification is impossible.
Any two elementary properties of a classical system
mutually exclude each other. However, this does not
hold anymore for a quantum system, where 
mutual exclusivity becomes a special relation
(orthogonality). This observation motivates
our mathematical model.

An orthogonality relation (axiom 1)
on some set is sufficient
to define states. In this simple setting, we introduce
two further basic axioms, which mimic
the quantum mechanical correspondence between
the pure states and the rank-one projections.
Using Gleason's theorem \cite{gleason1957}
and its extensions \cite{bunce1985Jordan, bunce1992mackey, Gleason-vector-valued}, 
we can identify a bunch of examples. They all
belong to a primary class, which results from certain operator
algebras and Jordan algebras \cite{AS02, hanche1984jordan}.
Whether further examples exist, remains an open issue.
Our mathematical model thus seems to possess more far-reaching
consequences than can be expected from its simplicity.

Moreover, we shall see that simple cross product
spaces do not result in a satisfying model for
multi-part systems and that the much larger tensor product is required,
which becomes responsible for the entanglement phenomenon
in multi-part quantum systems.

Furthermore we point out how our axioms give rise to some familiar 
mathematical structures that have often been used 
in earlier studies of quantum foundations and in the 
generalized probabilistic theories 
\cite{barnum2013ensemble, barrett2007information, muller2021probabilistic, plavala2023general}.
These are order-unit spaces \cite{AS02, hanche1984jordan}, 
quantum logics \cite{AS02, beltrametti2010logic}
and transition probability spaces \cite{mielnik1968geometry, pulmannova1986transition}.

Our approach is not restricted to the finite-dimensional
cases, but the non-atomic von Neumann algebras 
of type II and type III, which play a crucial role in quantum field theory,
are not included in our basic model. Finally we show
how it can be modified to encompass these cases.

The three basic axioms are introduced in the next section.
Some first examples satisfying them are investigated in section~3,
before the primary class is studied in section~4
and a few example candidates are discussed in section~5.
Bipartite systems and product spaces are considered
in section~6 and the connections to the other 
more familiar mathematical structures
in section~7. In section~8 we modify our model
in such a way that the non-atomic algebras are included.

\section{The basic axioms}

Let $E \neq \emptyset$ denote the set that consists of
the elementary properties of a system and let $\bot$ be a relation on $E$
satisfying\\

\noindent
\textbf{Axiom 1:} 
$ p \bot q \Rightarrow q \bot p$ \textit{and} $ p \neq q $\\

\noindent
for $p,q \in E$.
A family $F$ of elements in $E$ is called \emph{orthogonal},
if $p \bot q$ holds for any pair $p,q \in F$.
The \emph{capacity} $m$ of $E$ is the largest cardinality 
of the maximal orthogonal families. Note that the maximal
orthogonal families need not have the same size and that
their cardinality may be smaller than $m$.

The quantum mechanical superposition is already covered
by this simple model. The element $q \in E$ 
becomes the superposition 
of the elements $p_1, p_2 \in E$, when we have $p_1 \neq q \neq p_2$ and
$$e \in E, e \bot p_1, e \bot p_2 \Rightarrow e \bot q.$$

A \emph{state} $\mu$ is a map from $E$ to the real unit interval with 
$ \sum_{p \in F} \mu(p) = 1$
for any maximal orthogonal family $F$ in $E$.
When $F$ is not finite, the above sum is defined as the 
supremum of all finite partial sums. Each such state
becomes completely additive on each orthogonal family.
Let $S$ denote the space consisting of the states; $S$ is
a convex set.\\
\newpage
\noindent
\textbf{Axiom 2:}
\textit{For each $e \in E$, there is one and only one state $\mathbb{P}_e$ 
in $S$ with} $\mathbb{P}_e(e)=1$.\\

\noindent
Then $\mathbb{P}_e$ becomes an extreme point in $S$.\\

\noindent
\textbf{Axiom 3:}
\textit{For each extreme point $\mu$ in $S$, there is one and only one 
element $e$ in $E$ with $ \mu = \mathbb{P}_e $.}\\

The extreme points in $S$ are called \emph{pure states} 
and with the axioms 2 and 3 we obtain a one-to-one correspondence
between the pure states and the elements of $E$, which is similar
to the quantum mechanical correspondence between the pure states
and the rank-one projections.

The \emph{generalized probabilistic theories}
\cite{barnum2013ensemble, barrett2007information, muller2021probabilistic, plavala2023general}
usually start from a compact convex set that
represents the state space. Its extreme points
become the pure states, but it is not obvious
what should represent our set $E$ and how to
achieve the axioms 2 and 3. An approach in this
direction is presented in~\cite{nie2023conv_self-dual}.

\section{First examples}

We start with the consideration of the trivial 
relation $p \bot q \Leftrightarrow p \neq q$ on some set $E$.
With a state $\mu \in S$, the inequality $\mu(p) > 0$ 
holds for countably many $p$ at most and 
$S$ consists of the discrete probability measures on $E$. 
The measures concentrated in a single point are the pure states.
The three axioms are satisfied in this classical probability model. 
Superpositions do not exist here.\\

\noindent
\textbf{Lemma 1:}
\textit{If the axioms 1, 2 and 3 hold and $E$ contains a maximal orthogonal family 
with two elements $p,p'$, then $E = \left\{p,p'\right\}$}.\\

\textit{Proof}. Suppose that there is an element $q \in E$
beyond the maximal orthogonal family $p,p'$.
If $ q \bot p $, then $\mathbb{P}_q(p) = 0$, 
$\mathbb{P}_q(p') = 1$, $\mathbb{P}_q = \mathbb{P}_{p'}$ 
by axiom 1 and $ q = p' $ by axiom 2. In the same way we get that
$q = p$, if $q \bot p'$. Therefore $q$ is neither orthogonal 
to $p$ nor to $p'$. We select a $q_o$ with $p \neq q_o \neq p'$ 
and can then define a state $\mu_o$ by
$\mu_o(p) := 1 $, $\mu_o(p') := 0 $ and  
$\mu(q) := \mathbb{P}_{q_o}(q)$ for all other $q \in E$.
Then $\mathbb{P}_p = \mu_o = \mathbb{P}_{q_o}$ by axiom 1
and $p=q_o$ by axiom 2.\hfill $\square$\\

Lemma 1 rules out the triangular pillow \cite{AS02}
and all the qubit models \cite{Nie2022genqubit}.
With capacity $m=2$, there remains only one single case; 
this is the one with the trivial orthogonality relation
on a set $E$ with two elements.
Having this in mind, one might not expect anymore
that meaningful examples exist, but
our next example with capacity $m=3$ refutes that. 

Let $E$ consist of the
lines through the origin in three-dimensional Euclidean space
and let the relation $\bot$ on $E$ coincide with the usual orthogonality
of lines. The lines can be identified with the corresponding projection
operators or 3$\times$3-matrices $p$ and, by Gleason's theorem \cite{gleason1957}, 
the states $\mu$ on $E$ are the same as the positive semi-definite matrices
$\rho$ with trace 1. For each $\mu \in S$ there is a unique $\rho$
with $\mu(p) = trace(\rho p)$ for all projections $p$.
The axioms 2 and 3 can easily be verified; we shall reconsider this 
in the proof of our general theorem in the next section.
When the normalized vectors $\vec{x}$ and $\vec{y}$ 
generate the lines belonging to $p$ and $q$,
we we get $\mathbb{P}_p(q) = \left|\left\langle \vec{x}|\vec{y}\right\rangle\right|^{2}$,
which is the familiar form of the usual quantum mechanical transition probability.

In the same manner, the one-dimensional linear subspaces 
in any Euclidean space or complex inner product space with dimension $n \geq 3$
or the corresponding orthogonal projectors form a set $E$
which, with the usual orthogonality relation, satisfies the
axioms 1, 2, 3 and has the capacity $m=n$. 
The dimension $n=2$ must be excluded; one reason is Lemma~1,
and another one is that Gleason's theorem does not apply then.

Moreover, the one-dimensional linear subspaces
in any infinite-dimensional Hilbert space
or the atoms (minimal non-zero projections)
in any atomic von Neumann algebra 
without type $I_2$ part
satisfy our axioms
together with the usual orthogonality relation. 
The reason why the type $I_2$ parts must be ruled out
lies again in Lemma 1 and Gleason's theorem.

A further more exotic example are the points 
in the Moufang plane \cite{baez2002octonions}.
All these examples are covered by the atomic 
JBW-algebras \cite{AS02, hanche1984jordan}
which we will consider in the next section.

\section{The primary class of examples}

The JBW-algebras \cite{AS02, hanche1984jordan} are the Jordan analogue of the W*-algebras
and these represent the abstract axiomatization of the 
von Neumann algebras. A JBW-algebra $A$ possesses a commutative 
generally non-associative
product $\circ$ satisfying the Jordan identity 
$a^{2} \circ (a \circ b) = a \circ (a^{2} \circ b)$ ($a,b \in A$).
Moreover, $A$ possesses a multiplicative unit, which we denote by
$\mathbb{I}$, and an order relation~$\leq$; the positive cone is
$A_+ := \left\{a \in A | 0 \leq a\right\} = \left\{a^{2} | a \in A\right\}$.
The self-adjoint part~$A$ of a von Neumann algbra forms a JBW-algebra with
the product $a \circ b = $ \linebreak $(ab + ba)/2$.

Two idempotent elements $p$ and $q$ in a JBW-algebra $A$ are called \emph{orthogonal}
if $p \circ q = 0$. An atom is a non-zero idempotent element that 
cannot be written as the sum of two orthogonal non-zero idempotent elements.
The JBW-algebra $A$ is called \emph{atomic}, if
each non-zero idempotent element is the orthogonal sum of an atom 
and a further idempotent element. Then there is a non-negative 
trace function $trace: A_+ \rightarrow \left[0,\infty\right] $
with $trace(e)=1$ for every atom $e$. Moreover, 
$trace(a \circ b) \in \left[0,\infty\right]$ is well-defined for $a,b \in A_+$
and $trace(a \circ b)=0$ iff $a = 0 = b$~\cite{AS02}.

A JBW-algebra $A$ possesses a predual $A_\ast$,
which induces the weak topology $\tau(A,A_\ast)$ on $A$.
The linear functionals that are continuous with respect 
to this topology are called \textit{normal} and are identical
with those that are completely additive on orthogonal families
of projections. With an atomic algebra, the complete additivity
on orthogonal families of atoms becomes sufficient. This holds for the 
JBW-algebras in the same way as for the von Neumann algebras
or W*-algebras \cite{sakai1971}.
\newpage 

The finite-dimensional JBW-algebras coincide with the 
Euclidean (formally real) Jordan algebras. The irreducible cases
are the spin factors ($=$ type $I_2$ factors)
and the Jordan algebras that consist of the self-adjoint (Hermitian)
$m$$\times$$m$-matrices ($m \in \mathbb{N}$)
over the real and complex numbers, the quaternions and, only in the case $m=3$, 
the octonions \cite{AS02, hanche1984jordan}. These matrix algebras are denoted by 
$H_m(K)$ with $K = \mathbb{R},\mathbb{C},\mathbb{H},\mathbb{O}$.
The atoms in these matrix algebras form projective spaces;
with $H_3(\mathbb{O})$ this is the Moufang plane.\\

\noindent
\textbf{Theorem 1:} \textit{Let $E$ consist of the atoms in 
any atomic JBW-algebra $A$ without type $I_2$ part.
With the relation $p \bot q :\Leftrightarrow p\circ q=0$ ($p,q \in E$),
the axioms 1, 2 and 3 are satisfied. Moreover
$\mathbb{P}_q(p) = trace(q \circ p)$ for $p,q \in E$ and,
if $\mathbb{P}_p(q) = 0$, then $p \bot q$.}\\

\textit{Proof.} Axiom 1 is obvious. By a generalization of Gleason's theorem
\cite{bunce1985Jordan, bunce1992mackey, Gleason-vector-valued},
the states on $E$ are identical with the restrictions of those
positive linear functionals that are completely additive 
on orthogonal families of atoms. These are identical with the normal functionals 
and thus have the form (see \cite{AS02} Proposition 5.66)
$$ E \ni p \rightarrow \mu(p) = \sum^{\infty}_{k=1} s_k trace(e_k \circ p)$$
with an orthogonal family of atoms $e_k$, $0 \leq s_k \leq 1$ and
$\sum^{\infty}_{k=1} s_k = 1$. Here $\sum s_k e_k$ is a so-called 
trace class element in $A$.

If $\mu(q)=1$ with $q \in E $, then $trace (e_k \circ q) = 1$ for each $k$ with $s_k \neq 0$.
However $trace (e_k \circ q) = 1$ can hold for the atoms $e_k$ and $q$ only if $e_k = q$. Therefore,
$q = e_{k_o}$ for one $k_o$, $s_k = 0$ for all $k \neq k_o$ and $s_{k_o} = 1$. This means 
$\mu(p) = trace(q \circ p)$ for $p \in E$ and we have axiom 2
with $\mathbb{P}_q(p) = trace(q \circ p)$. Furthermore,
if $0 = \mathbb{P}_p(q) = trace(p \circ q)$ with $p,q \in E$, then $p \circ q = 0$.

The pure states in $S$ are identical with the extremal positive trace class elements
with normalized trace in $A$ and these are the atoms. Thus we also have axiom 3.\hfill $\square$

\section{Example candidates}

The atoms in the matrix algebras $H_m(\mathbb{R})$, $H_m(\mathbb{C})$, $H_m(\mathbb{H})$
with $m \geq 3$ and $H_3(\mathbb{O})$ form the projective spaces 
$\mathbb{R}\mathbb{P}^{n}$, $\mathbb{C}\mathbb{P}^{n}$, 
$\mathbb{H}\mathbb{P}^{n}$ ($n \geq 2$) and $\mathbb{O}\mathbb{P}^{2}$
\cite{baez2002octonions}. These are
compact connected Riemannian manifolds, 
where the geodesic distance between two points $p$ and $q$ becomes
$arccos (trace^{1/2} (p\circ q ))$.
We thus get that $p$ and $q$ are orthogonal 
($p \bot q \Leftrightarrow p \circ q = 0 \Leftrightarrow trace(p \circ q) = 0$)
iff the points $p$ and $q$ achieve the maximal geodesic distance
$\pi /2$ in these manifolds.
Therefore the question arises, whether further compact connected 
Riemannian manifolds satisfy our axioms 1, 2 and 3 with the definition 
that two points become orthogonal iff they achieve the maximal geodesic distance.

First candidates might be the generalized projective planes
$\left(\mathbb{C} \otimes \mathbb{O}\right)\mathbb{P}^{2}$, \linebreak
$\left(\mathbb{H} \otimes \mathbb{O}\right)\mathbb{P}^{2}$ 
and $\left(\mathbb{O} \otimes \mathbb{O}\right)\mathbb{P}^{2}$~\cite{baez2002octonions}.
The isometry groups in these three cases are the exceptional Lie groups 
$E_6$, $E_7$ and~$E_8$.
Interestingly, $E_6$ is sometimes considered a candidate for internal symmetries
in particle physics~\cite{catto2003, gursey1978octonionic};
$E_7$ and $E_8$ are candidates for unifying the fundamental forces and particles
in the \emph{Grand Unified Theories} (GUTs) 
and \emph{String Theory} \cite{heckman2010particle, KUGO1984}.

However, the verification or falsification of our axioms for these cases remains an 
open issue. Note that a kind of Gleason theorem would be needed for the
verification.

\section{Product spaces}

When each one of the sets $E_1$ and $E_2$
is equipped with an orthogonality relation~$\bot$,
we can easily define an orthogonality relation on the
product set $E_1 \times E_2$ in the following way:
\begin{center}
$(p_1,p_2) \bot (q_1,q_2) \  :\Leftrightarrow \  p_1 \bot q_1$ \textit{or} $p_2 \bot q_2$
\end{center}
for $p_1, q_1 \in E_1$ and  $p_2, q_2 \in E_2$. However, our axioms 
do not remain valid for $E_1 \times E_2$, when they hold in both $E_1$ and $E_2$.
We cannot demonstrate this in the general case, but in the familiar situation
of finite-dimensional common quantum theory. By $E_m(\mathbb{C})$ we denote
the system of the atoms in the matrix algebra $H_m(\mathbb{C})$, $m \in \mathbb{N}$.\\

\noindent
\textbf{Lemma 2:} $E_m(\mathbb{C}) \times E_n(\mathbb{C})$ \textit{violates axiom} 3.\\

\textit{Proof.}  Let $\mu$ be a state on $E_m(\mathbb{C}) \times E_n(\mathbb{C})$.
When $p \in E_m(\mathbb{C})$ is fixed, we apply the Gleason theorem and extend the map 
$E_n \ni q \rightarrow \mu(p \times q)$ on $E_n$
to a linear map on the Jordan matrix algebra $H_n(\mathbb{C})$.
With a fixed element $b \in H_n(\mathbb{C})$ we can then extend
the map $E_m \ni p \rightarrow \mu(p \times b)$ on $E_m$
to a bilinear map on $H_{m}(\mathbb{C}) \times H_{n}(\mathbb{C})$.
Using the Kronecker product $\otimes$ for matrices, we can embed 
$H_{m}(\mathbb{C}) \times H_{n}(\mathbb{C})$ in $H_{mn}(\mathbb{C})$,
where the bilinear map on $H_{m}(\mathbb{C}) \times H_{n}(\mathbb{C})$
has a linear extension to the linear tensor product
$H_{m}(\mathbb{C}) \otimes H_{n}(\mathbb{C}) \subseteq H_{mn}(\mathbb{C})$.
Since both have the same real dimension $m^{2}n^{2}$, we have
$H_{m}(\mathbb{C}) \otimes H_{m}(\mathbb{C}) = H_{mn}(\mathbb{C})$.
All these map extensions are unique, but it is well-known that
they are not always positive \cite{barnum2005influencefree, Groe1985,
klay1987tensor}. This means that the state space of
$E_{mn}(\mathbb{C})$ is a proper subset of the state space of
$ E_m(\mathbb{C}) \times E_n(\mathbb{C}) $. 

Both are the closed convex hulls of their extreme points.
If axiom 3 would hold for $ E_m(\mathbb{C}) \times E_n(\mathbb{C}) $,
its state space would be the closed convex hull of 
$\left\{\mathbb{P}_{e_1 \otimes e_1} | e_1 \in E_m(\mathbb{C}), e_2 \in E_n(\mathbb{C})\right\}$
and the two state spaces would become identical.\hfill $\square$\\

This shows that the simple cross product $E_m(\mathbb{C}) \times E_n(\mathbb{C})$
is not sufficient to model a composite system, but that
the much larger space $E_{mn}(\mathbb{C})$ is required.
An important consequence is the entanglement phenomenon in multi-part
quantum systems. The model for a $n$-qubit system is $E_{2^{n}}(\mathbb{C})$
and satisfies our axioms for~$n > 1$. The single isolated
qubit is ruled out by Lemma 1.

\section{Order-unit spaces, quantum logics and transition probability spaces}

In this section we show how our basic model gives rise to some familiar 
mathematical structures that have often been used in the 
generalized probabilistic theories 
\cite{barnum2013ensemble, barrett2007information, muller2021probabilistic, plavala2023general}
and in the earlier studies of quantum foundations. These are
order-unit spaces \cite{AS02, hanche1984jordan}, quantum logics \cite{AS02}
and transition probability spaces~\cite{mielnik1968geometry, pulmannova1986transition}.

Let $E$ be a set with a relation $\bot$,
which satisfies the axioms 1, 2 and 3.
By $A_b(S)$ we denote the space that consists
of the bounded real-valued affine functions on the 
state space $S$. Its elements represent 
the (bounded) observables in the same way 
as in the generalized probabilistic theories
\cite{barnum2013ensemble, barrett2007information, muller2021probabilistic, plavala2023general}.
With the supremum norm $\left\|\ \right\|$,
$A_b(S)$ becomes a complete
order-unit space \cite{AS01, hanche1984jordan},
where $a \leq b$ iff $a(\mu) \leq b(\mu)$ for all $\mu \in S$.
The order-unit is the constant function 
with the value $1$ and is denoted by $\mathbb{I}$.
The space $A_b(S)$ is also complete with respect to the point-wise
convergence on~$S$. Each $\mu \in S$ results in the 
state $ \bar{\mu} :a \rightarrow a(\mu) $ on $A_b(S)$.
We do not formally distinguish between $\mu$ and $\bar{\mu}$
and write $ a(\mu) = \mu(a) $.

For each $e \in E$ we define an element $\bar{e}$ in $A_b(S)$ by
$\bar{e}(\mu) := \mu(e)$ for $\mu \in S$. Then $0 \leq \bar{e} \leq \mathbb{I}$.
The identity $\bar{e_1} = \bar{e_2}$ for $e_1, e_2 \in E$ implies
$\mu(e_1) = \mu(e_2)$ for all $\mu \in S$ and particularly
for $\mu = \mathbb{P}_{e_1}$. We thus get 
$1 = \mathbb{P}_{e_1}(e_1) = \mathbb{P}_{e_1}(e_2)$.
Then $\mathbb{P}_{e_1} = \mathbb{P}_{e_2}$ by axiom 2
and finally $e_1 = e_2$ by axiom 3. This means that
the map $e \rightarrow \bar{e}$ becomes an embedding
of $E$ in 
$\left[0,\mathbb{I}\right] 
:= \left\{a \in A_b(S) | 0 \leq a \leq \mathbb{I}\right\} \subseteq A_b(S)$.
For any orthogonal family $J$ in $E$ we get the following element
$a_J$ in $\left[0,\mathbb{I}\right]$:
$$a_J := \lim\limits_{I \in J_f} \sum_{e \in I} \bar{e},$$
where $J_f$ consists of all finite subsets of $J$
and the generalized sequence (net) $a_I = \Sigma_{e \in I} \bar{e}$,
$I \in J_F$, converges pointwise on $S$.
From now on we simply write $a_J = \sum_{e \in J} \bar{e}$.
For each maximal $J$ we have $a_J = \mathbb{I}$. Moreover $a_{\emptyset}=0$. 

Now let $L$ denote the set of those elements $a$ in $A_b(S)$, 
for which there is an orthogonal family $J$ in $E$ with $a = a_J$.
The smallest element in $L$ is $0$ and the largest one is
$\mathbb{I}$. With the orthocomplementation $a' := \mathbb{I} - a$, 
$L$ becomes an excellent candidate for a quantum logic \cite{AS02, beltrametti2010logic}.
We have $a' \in L$, since any orthogonal family $J$ with $a=a_J$
can be enlarged to a maximal orthogonal family $I$ and then
$a' = a_{I - J}$.

With $p(x,y):=\mathbb{P}_x(y)$ for $x,y \in E$,
we get a \emph{transition probability space},
such as B. Mielnik~\cite{mielnik1968geometry} defined it,
when we furthermore postulate $\mathbb{P}_x(y) = \mathbb{P}_y(x)$.
S.~Pulmannova~\cite{pulmannova1986transition} considered 
transition probability spaces without this symmetry postulate
and derived some atomic quantum logical structures. 
Vice versa, an atomic
quantum logic has an orthogonality relation 
and we can select the atoms 
(minimal non-zero elements) to form the set $E$,
but the axioms~2 and~3 need not hold then and have 
to be postulated in addition.
\newpage

In the transition probability spaces,
the orthogonality relation is defined in the following way: 
$p \bot y :\Leftrightarrow p(x,y) = 0$. It coincides 
with our one in the atomic JBW-algebras (Theorem 1),
but it does so in our general setting
only if the following further axiom is satisfied.\\

\noindent
\textbf{Axiom 4:} \textit{If $\mathbb{P}_p(q) = 0$ holds with $p, q \in E$, then $p \bot q$}.\\

A pair $a,b \in L$ is called \emph{orthogonal} and we write 
$a \bot b$, if $a + b \leq \mathbb{I}$.
We always have $ \bar{p} \bot \bar{q} $ in $L$ for $p,q \in E$ with $p \bot q$ in $E$.
With axiom 4, we also have that $p \bot q$ holds for $p,q \in E$ 
with $ \bar{p} \bot \bar{q} $ in $L$, since the inequality
$ \bar{p} + \bar{q} \leq \mathbb{I}$ means 
$\mu(p) + \mu(q) \leq 1$ for all $\mu  \in S$ 
and particularly for $\mu = \mathbb{P}_p$,
thus $1 + \mathbb{P}_p(q) \leq 1$ and $\mathbb{P}_p(q)=0$.

Generally, $L$ is not a lattice, since the supremum $\vee$
and the infimum $\wedge$ do not always exist. 
The supremum does exist for pairwise orthogonal $b_k$ in $L$.
To see this, select orthogonal families $J_k$ with $b_k = a_{J_k}$. Then
$I := \cup_k \  J_k$ becomes an orthogonal family.
Obviously $b_k \leq a_I = \sum_k b_k $ for each $k$. To see that $a_I$
is the smallest upper bound, suppose $b_k \leq b$ for each $k$
with $b = a_K $, $\mathbb{I} - b = a_{K'} $ and 
orthogonal families $K,K' \subseteq E$.
Then $I \cup K'$ is an orthogonal family and thus 
$ a_I + \mathbb{I} - b = a_I + a_{K'} \leq \mathbb{I}$.
Therefore $a_I \leq b$.\\

\noindent
\textbf{Lemma 3:} \textit{We assume that axiom 4 holds.}

\noindent
(i) $a_1, a_2 \in L, a_1 \leq a_2 \Rightarrow a_2 = a_1 \vee (a_1' \wedge a_2)$.

\noindent
(ii) $a,b \in L$ \textit{and} 
$ \left\{\mu \in S | \mu(a)=1\right\} \subseteq \left\{\mu \in S | \mu(b)=1\right\} 
\Rightarrow a \leq b$.\\

\textit{Proof}. (i) Suppose $a_1 \leq a_2$ with $a_1, a_2 \in L$.
Then there are orthogonal families $J_1, J_2, J_3$ in $E$ with
$a_1 = a_{J_1}$, $a_2 = a_{J_2}$, $a_2' = a_{J_3}$.
If $e_1 \in J_1$ and $e_3 \in J_3$, then 
$\bar{e_1} + \bar{e_3} \leq a_1 + a_2' \leq a_2 + a_2'= \mathbb{I}$
and $e_1 \bot e_2$ (see above).
Thus $J_1 \cup J_3$ is an orthogonal family and can be enlarged
to a maximal orthogonal family $I$. 
With $J_4 := I - (J_1 \cup F_3)$ we get
$\mathbb{I} = a_{J_1} + a_{J_3} + a_{J_4}
= a_1 + a_2' + a_{J_4}$ and
$a_2 = \mathbb{I} - a_2' = a_1 + a_{J_4} = a_1 \vee a_{J_4}$ (see above). 
Moreover, $a_1 \vee a_2' = a_{J_4}'$ and thus
$a_{J_4} = a_1' \wedge a_2$.

(ii) Suppose $\left\{\mu \in S | \mu(a)=1\right\} \subseteq \left\{\mu \in S | \mu(b)=1\right\}$
with $a,b \in L$, $a = a_{J_1}$, $b' = a_{J_2}$ with the orthogonal families $J_1$ and $J_2$ in $E$.
Then $\mathbb{P}_{e_1}(b) = 1$ and $\mathbb{P}_{e_1}(b') = 0$ for each $e_1 \in J_1$. 
Thus $\mathbb{P}_{e_1}(e_2) = 0$ and $e_1 \bot e_2$ for each $e_1 \in J_1$ and each $e_2 \in J_2$.
Therefore $a \bot b'$ and $a \leq b$.\hfill $\square$ \\

Part (i) of Lemma 3 is the \emph{orthomodular law} \cite{AS02, hanche1984jordan} 
and part (ii) means that the state space $S$ is 
\emph{strongly order determining} \cite{nie2023conv_self-dual}.

Our quantum logic $L$ is isomorphic to one of those which
S. Pulmannova~\cite{pulmannova1986transition} constructed 
(the one formed by the objects she calls "orthoclosed subsets").

Our states on $E$ are the same as the completely orthogonally additive
non-negative measures $\mu$ on $L$ with $\mu(\mathbb{I}) = 1$ and, 
in our setting, each such measure possesses a linear extension on $A_b(S)$.
This means that we get a kind of Gleason theorem, here in a rather trivial way, 
from our axioms. Recall that we have needed this theorem
for the verification of the axioms in all examples
with non-trivial orthogonality relations.

\section{Non-atomic structures}

In quantum field theory the infinite-dimensional 
non-atomic von Neumann algebras of type II and type III
play a crucial role. Mielnik's transition probability space framework
\cite{mielnik1968geometry}
as well as our approach so far are not restricted to the finite-dimensional cases, 
but these non-atomic cases are not included. 
We shall now show how our approach can  be 
modified in such a way that they are included.
The non-existence of atoms
means that there are no \emph{elementary} system properties.
Therefore we cannot start with these and must
consider the set that consists of \emph{all} system properties,
which makes the situation more tricky.

Let $L_o$ be any non-empty set with a relation $\bot$
that satisfies axiom 1 and define the state space $S$ as earlier.
The following simple axiom shall ensure that $S$ becomes 
point-separating on $L_o$. In section~2 the pure states were
point-separating on $E$, which followed from the axioms~2 and 3.\\

\noindent
\textbf{Axiom A:} \textit{For every pair $p,q \in L_o$ with $p \neq q$ 
there is a state $\mu \in S$ with $\mu(p) \neq \mu(q)$}.\\

By $A_b(S)$ we again denote the order-unit space that consists
of the bounded real-valued affine functions on the 
state space $S$, and for each $p \in L_o$ we define an element $\bar{p}$ in $A_b(S)$ by
$\bar{p}(\mu) := \mu(p)$ for $\mu \in S$. 
The identity $\bar{p} = \bar{q}$ for $p,q \in L_o$ implies
$\mu(p) = \mu(q)$ for all $\mu \in S$ and thus $p = q$ by axiom~A.
This means that the map $p \rightarrow \bar{p}$ again becomes an embedding
of $L_o$ in $\left[0,\mathbb{I}\right] 
:= \left\{a \in A_b(S) | 0 \leq a \leq \mathbb{I}\right\} \subseteq A_b(S)$.
As in section 4 let $L$ denote the set of those elements $a$ in $A_b(S)$, 
for which there is an orthogonal family $J$ in $L_o$ with $a = a_J$.
The smallest element in $L$ is $0$, the largest one is
$\mathbb{I}$ and, with the orthocomplementation $a' := \mathbb{I} - a$, 
we have $a' \in L$, since any orthogonal family $J$ with $a=a_J$
can be enlarged to a maximal orthogonal family $I$ and then
$a' = a_{I - J}$. We have $\bar{p} + \bar{q} \leq \mathbb{I}$,
if $p,q \in L_o$ and $p \bot q$. With the following axiom we get
$\bar{p} + \bar{q} \leq \mathbb{I}$ iff $p \bot q$.\\

\noindent
\textbf{Axiom B:} \textit{If the inclusion 
$ \left\{\mu \in S | \mu(p) = 1\right\} \subseteq \left\{\mu \in S | \mu(q) = 0\right\}$ 
holds for $p,q \in L_o$, then $ p \bot q $}.\\

This is a strengthened version of axiom 4.
Moreover, axiom B means that the state space~$S$
is strongly \emph{orthogonality} determining
in a similar way as it is strongly \emph{order} determining
in Lemma 3 (ii). Furthermore, for each $p \in L_o$ there is
a state $\mu \in S$ with $\mu(p) = 1$; if there were
no such state, axiom B would imply $p \bot p$, 
which is impossible by axiom 1.

Since we cannot assume that $S$ contains extreme points,
we have to consider the faces of $S$. Moreover,
when the dimension is not finite, a topology is usually needed,
and we equip the state space $S$ with the following metric:

$$d(\mu_1,\mu_2) := sup \left\{\left|\mu_1(p) - \mu_2(p) \right| : p \in L_o \right\}$$
for $\mu_1, \mu_2 \in S$.
A convex subset $F$ of $S$ is called a \emph{face} if
\begin{center}
$\mu_1, \mu_2 \in S$, $0 < s < 1$, 
$s \mu_1 + (1-s) \mu_2 \in F$ $ 
\Rightarrow$ $\mu_1, \mu_2 \in F$
\end{center}
holds. If the singleton $\left\{\mu\right\}$ with $\mu \in S$ is a face,
$\mu$ becomes an extreme point of $S$.
With the usual set inclusion as ordering,
the faces and those that are closed with respect to the metric $d$
form complete lattices, always 
containing the elements $\emptyset$ and $S$.
For each $p\in L_o$ we have the $d$-closed face
$$F_p := \left\{\mu \in S | \mu(p)=1\right\}.$$
Obviously we have $F_p \subseteq F_q$, 
if $\bar{p} \leq \bar{q}$ with $p,q \in L_o$.
As we shall see now, with axiom B we also get 
$\bar{p} \leq \bar{q}$, 
if $F_p \subseteq F_q$. This means that 
the state space $S$ becomes again strongly order determining 
for $L$ and, furthermore, $F_p = F_q$ iff $\bar{p} = \bar{q}$,
which is equivalent to $p = q$.

Suppose $p,q \in L_o$ with $F_p \subseteq F_q$ and
let $J$ be an orthogonal family in $L_o $ with $a_J = \mathbb{I} - \bar{q}$. 
With any $\mu \in S$ we then have
$$\mu(p)=1 \Rightarrow \mu(q) =1 \Rightarrow \mu(a_J)=0 \Rightarrow \mu(e) = 0 $$
for each $ e \in J $ and, by axiom B, $p \bot e$ for each $ e \in J $. 
Thus $ \left\{ p \right\} \cup J $ becomes an orthogonal family in $L_o$.
Therefore $ \bar{p} + \mathbb{I} - \bar{q} = \bar{p} + a_J \leq \mathbb{I} $
and $\bar{p} \leq \bar{q} $.

After these preparations we can now introduce an axiom, 
which on the one hand is a strengthened version of axiom 3,
but on the other hand encompasses the non-atomic cases.\\

\noindent
\textbf{Axiom C:} \textit{For each non-empty $d$-closed face $F$ of $S$, 
there is an element $p \in L_o$ with $F = F_p$}.\\

This axiom may have something in common with Alfsen and Shultz's concept of 
projective faces \cite{AS02}, but they require that $p = Q \mathbb{I}$ 
with a \emph{compression} $Q$. These compressions are maps with some
very special properties and are not available here.

With axiom C the map
$\bar{p} \rightarrow p \rightarrow F_p$ becomes
an order preserving bijection between $L \setminus \left\{0\right\}$
and the non-empty $d$-closed faces of the state space $S$.
Hence $L$ is a complete lattice, which is more than we had earlier. 
It is orthomodular with the orthocomplementation $a' := \mathbb{I} - a$,
which can be shown in the same was as 
in the proof of part (i) of Lemma~3.
Moreover, $L \setminus \left\{0\right\} $ 
is identical with the embedding of $L_o$.

The state space $S$ is convex, but need not be compact. Therefore
it need not contain extreme points. If it does,
the faces belonging to these pure states 
are singletons. The elements they contain 
result in atoms in the quantum logic~$L$ and
can be used to form the set $E$, considered earlier in this paper.
If $E$ is sufficiently large, 
the axioms 2, 3 and 4 are satisfied
(for instance, if each non-empty closed face of $S$
contains a pure state or, equivalently, 
if the quantum logic is atomic and, particularly,
if $S$ is the convex hull of the pure states
or finite-dimensional).
\newpage

After removing the zero-element, the axioms A and B hold 
in any orthomodular partially ordered set
(quantum logic), on that the completely additive states are 
separating and strongly order determining.
Axiom C becomes the new ingredient here. Moreover, it is not
necessary to start with a quantum logic; a set with an
orthogonality relation is sufficient to define the state 
space, with which we can proceed then.

The Borel sets in $\mathbb{R}$ form a $\sigma$-algebra
and, after removing the element $\emptyset$, we get a system $L_o$,
where two elements are orthogonal iff they are disjoint.
The state space $S$ consists of the discrete probability measures 
on $\mathbb{R}$ and $A_b(S)$ is isomorphic to 
the algebra of all bounded functions on $\mathbb{R}$.
This is an atomic associative JBW-algebra. Then
$L$ is the power set of $\mathbb{R}$. Axiom C does not hold here,
but becomes valid when we replace the above $L_o$ 
by the the power set of $\mathbb{R}$.

We can use the Borel sets and then form the quotient space 
with the Lebesgue null sets. After again removing the zero-element,
we get another space $L_o$,
where the state space consists of the Lebesgue-continuous classical
probability measures and $A_b(S) \cong L^{\infty}$. This is
another associative JBW-algebra, but it possesses neither atoms nor 
normal pure states. Here axiom C is satisfied.

The non-zero projections in any JBW-algebra $A$ without type $I_2$ part 
form a system $L_o$ that satisfies
the axioms A, B, C with the order relation 
$p \bot q :\Leftrightarrow p \circ q = 0$, $p,q \in L_o$. 
Here $S$ consists of the normal 
(completely additive) states on $A$, since the Gleason theorem 
\cite{gleason1957} and its extensions 
\cite{bunce1985Jordan, bunce1992mackey, Gleason-vector-valued}
again ensure that our states have linear extensions to states on $A$.
The axioms B and C follow from~\cite{AS02} Theorem~5.32; 
note that $d$ and the metric arising from the usual norm
are equivalent, since the unit ball in $A$ is the closed convex hull 
of $L_o \cup \left\{-p | p \in L_o\right\}$, due to the spectral theorem.

The structures of the state space $S$, the quantum logic $L$ and
the system of observables $A_b(S)$ are completely hidden in the orthogonality 
relation $\bot$, but the characteristics of the resulting
structures are not immediately visible. The quantum logic 
may turn out to be Boolean (i.e. classical) or atomic or non-atomic and
an emerging JBW algebra may be associative, atomic, of type II, III or
a mixture.

\section{Conclusions}

Starting from the observation that 
the distinction between inequality and mutual exclusivity 
of elementary system properties distinguishes
quantum theory from classical theories,
we have developed a simple mathematical model.
It consists of nothing more than
an orthogonality relation on some set,
which is enough to define states, and two
postulates for the states, which mimic
the quantum mechanical correspondence between
the pure states and the rank-one projections.
We have investigated a bunch of examples and, interestingly,
they all origin from operator algebras and certain Jordan algebras.
The identification of further cases remains an unresolved challenge.
Note that a kind of Gleason theorem would be required.
If further cases exist, they might result in promising
mathematical models for an extension of quantum theory. The non-existence
would mean that our basic postulates bring us unexpectedly close to
common quantum theory. Thus our model seems to possess more far-reaching
consequences than can be expected from its simplicity.

In any case, the fact that inequality and
mutual exclusivity of elementary system properties
are not the same in quantum physics appears to be a major origin
of quantumness. The superposition phenomenon arises immediately 
and some further axiom for the state space results
in the necessity of entanglement in multi-part systems.

Furthermore, we have pointed out how our model can be modified 
to encompass the non-atomic von Neumann algebras of type II, III,
which play a crucial role in quantum field theory, but where 
rank-one projections (atoms) which represent the elementary system 
properties are not available.

A major ingredient in our approach is the orthogonality relation. Though
this is a rather simple thing at first glance,
it can conceal a large amount of mathematical structure,
which has been known since A. Dye showed in 1955 that any
orthoisomorphism between the projections in two 
von Neumann algebras (without type $I_2$ part)
possesses a linear extension which becomes a Jordan *-isomorphism \cite{dye1955geo}.
Different extensions of Dye's result have been achieved since
then \cite{bunce1993dye, Hamhalter2015Dye, hamhalter2021dye}

\bibliographystyle{abbrv}
\bibliography{Literatur2026}
\end{document}